\documentclass[10pt,aps,prl,twocolumn,preprintnumbers,amsmath,amssymb,floatfix,showpacs,citeautoscript]{revtex4-1}
\usepackage[latin1]{inputenc}
\usepackage[T1]{fontenc}
\usepackage[english]{babel}
\usepackage[pdftex]{graphicx}
\usepackage{amsmath}
\usepackage{times}
\usepackage[usenames,dvipsnames,svgnames,table]{xcolor}
\usepackage[colorlinks,plainpages=false,linkcolor=blue,urlcolor=blue,citecolor=blue,pdfpagemode=UseNone]{hyperref}

\renewcommand{\AA}{\text{\r{A}}}

\begin{document}

\title
{
\boldmath
Adsorption and dissociation of iron phthalocyanine on H/Si$(111)$:
Impact of van-der-Waals interactions and perspectives for subsurface doping
}

\author{Benjamin Geisler}
\email{benjamin.geisler@uni-due.de}
\affiliation{Fakult\"at f\"ur Physik, Universit\"at Duisburg-Essen and Center for Nanointegration (CENIDE), Campus Duisburg, Lotharstr.~1, 47048 Duisburg, Germany}
\author{Peter Kratzer}
\affiliation{Fakult\"at f\"ur Physik, Universit\"at Duisburg-Essen and Center for Nanointegration (CENIDE), Campus Duisburg, Lotharstr.~1, 47048 Duisburg, Germany}

\date{\today}

\begin{abstract}
The adsorption of iron phthalocyanine (FePc) on the passivated H/Si$(111)$ surface is explored from first principles.
We find that the organic molecule is predominantly physisorbed with a distance to the surface of $2.6 \pm 0.1~\AA$ and an adsorption energy of $1.55 \pm 0.1$~eV,
but also exhibits sizable resonance with the underlying substrate.
This establishes the present system as interesting mixed covalent-van-der-Waals-bound test case,
which we use to compare the impact of different approaches to van-der-Waals interactions.
(Spin-polarized) scanning tunneling microscopy (SP~STM) images are simulated, selectively accessing different molecular orbitals via the applied bias voltage in the spirit of scanning tunneling spectroscopy.
Comparison with experimental STM images reveals very good agreement.
We report a significant magnetic contrast exceeding $\pm 1~\AA$ in the SP~STM images for $-2$ and $+1.5$~V.
Aiming for a magnetic functionalization of Si for possible spintronics applications,
magnetic moments and binding energies of different (transition metal) atoms in the center of the Pc ring are presented,
which particularly show that Fe is strongly bound in the molecule (about $9.6$~eV).
Finally, we discuss different mechanisms for subsurface Fe doping by room-temperature FePc deposition and point out two feasible reactions.
Concomitantly, we identify the crucial role of a preceding destabilization of FePc, for instance, by pre-adsorbed H atoms,
which subsequently strongly stabilize the final state of the reaction.
\end{abstract}

% \pacs{73.20.-r, 68.65.Cd, 71.30.+h, 74.78.Fk, 72.15.Jf}

\maketitle

\section{Introduction}

The exploration of viable routes towards Si-based spintronics devices~\cite{Jansen:12} is of key importance,
owing to the abundance of Si and its unmatched role in modern semiconductor technology.
Magnetic functionalization of Si by a controlled doping with transition metal (TM) impurities
is a potential strategy.
Despite intense research, comprising
recent studies of formation energies and magnetic properties of isolated TM impurities and impurity pairs~\cite{Zhang:08},
the dependence of magnetism on the doping concentration~\cite{KuewenBechstedt:09,Shaughnessy:10},
or Mn $\delta$ doping of Si~\cite{Qian:06, Wu:07, Otrokov:11},
at present Mn-doped GaAs dominates the field~\cite{Ohno-MnGaAs:96, KoenraadFlatte:11, Jancu:08}.

For a rational materials design,
it is crucial to improve the fundamental understanding of the interaction mechanisms
between impurities and the surrounding host matrix.
We recently promoted (spin-polarized) scanning tunneling microscopy (SP STM)~\cite{Wiesendanger-RevModPhys:09}
as a powerful method
to explore bulklike interaction properties of impurities in semiconductors on the atomic scale
by exploiting passivated surfaces~\cite{Geisler-TMs:15}.
Specifically, we focused on TM impurities below the H/Si$(111)$ surface.
The passivating H layer could be added after growth and subsequent cleavage of a sample;
however, exposing the doped system to a wet-chemical treatment~\cite{Gruyters:13} will cause strong reactions of the TM impurities with H.
Hence, a strategy to achieve a controlled subsurface doping of the H/Si$(111)$ system after preparation would be valuable.
It was claimed recently that this can be accomplished by the adsorption of iron phthalocyanine (FePc) on the H/Si$(111)$ surface~\cite{Gruyters:13}.
This is particularly compelling
owing to the fact that Fe is one of the fundamentally most interesting impurities in Si,
since its magnetic moment could not be unambiguously determined from first principles so far~\cite{Geisler-TMs:15}.

The study of surface reactions in hybrid inorganic-organic systems
is a rich topic of its own.
Considerable research focuses on the adsorption of molecules on
metallic~\cite{Morbec-AnthraPentaAg111:17, Herper-FePc-Co001:14, Ruiz:12},
semiconducting~\cite{Veiga-Mol-Si111:16, Gruyters:12, Sena:09},
or insulating~\cite{Repp:05}
surfaces,
magnetic properties of TM-based molecules and their interactions with ferro- and nonmagnetic substrates~\cite{Bernien:09, Atodiresei:10},
and magnetic switching phenomena~\cite{WendeGraphene:11}, which aim at spin-dependent molecular electronics~\cite{Wende:07, Kratzer-Porphyrins-Graphene:17}.
The Kondo effect can be observed in magnetic molecules deposited on metal surfaces~\cite{KondoMnPc:07, PereraKondo:10}.
Given that the impact of the surface on the electronic structure of the adsorbate is minute,
these systems are also attractive models to gain fundamental insight by comparing experimental and time-dependent density functional theory (TDDFT) results,
for instance, on optical properties of Pc-derived molecules~\cite{Cocchi-TDDFT-Pc:14}.
Naturally, van-der-Waals (vdW) interactions are of central relevance in this field.

Here we explore the adsorption of FePc on H/Si$(111)$ from first principles.
We find that the organic molecule is predominantly physisorbed,
but also shows significant resonance with the underlying substrate.
We use this paradigmatic case of a system in a mixed covalent-vdW-bonding state
to compare the impact of different established~\cite{Grimme:06, Grimme-D3:10} and recent~\cite{Tkatchenko:09, Tkatchenko:12} approaches to vdW interactions,
thereby going beyond small model systems that are conventionally used for optimizing and testing these dispersion methods.
Analogies to 2D materials, e.g., graphene, boron nitride, or transition metal dichalcogenides add to the relevance of such benchmarking~\cite{Tawfik:18}.
(SP)~STM images are simulated, selectively accessing different molecular orbitals via the applied bias voltage in the spirit of scanning tunneling spectroscopy,
and compared to experiment.
The SP~STM images exhibit a magnetic contrast that exceeds $\pm 1~\AA$ for $-2$ and $+1.5$~V.
In order to shed light on the mechanism behind the suggested subsurface doping involving room-temperature FePc deposition,
we present binding energies of different (TM) atoms in the center of the Pc ring,
which particularly reveal that Fe is strongly bound (about $9.6$~eV).
This result is confirmed by different methodologies, including hybrid functionals, and holds for other $3d$ TM centers as well.
Despite this surprising finding,
we identify two feasible reactions by analyzing several distinct mechanisms to achieve subsurface Fe doping.
Simultaneously, we highlight the
% crucial role of a preceding destabilization of FePc, for instance, by pre-adsorbed H atoms.
% \textcolor{blue}{crucial role of stabilizing the final state of the reaction, which requires two H atoms to be provided by the initial reactant FePc.
% Moreover, FePc-2H (not known here!) is less stable than FePc alone and thus more prone to reaction.}
crucial role of a preceding destabilization of the reactant FePc, for instance, by pre-adsorbed H atoms,
which subsequently strongly stabilize the final state of the reaction.

\section{Methodology}

We performed spin-polarized density functional theory~\cite{KoSh65} (DFT) calculations
within the plane-wave and ultrasoft pseudopotential (PP) framework
as implemented in the Quantum Espresso code~\cite{Vanderbilt:1990, PWSCF}
and compare to results obtained with
the full-potential, all-electron (AE) FHI-aims code~\cite{FHIaims}
which employs numeric atom-centered basis functions.
We use cutoff energies for wave functions and density of $35$ and $250$~Ry
and a 'tight' 'tier2' basis, respectively.
Gas-phase molecules were simulated with different exchange-correlation functionals,
namely semilocal LDA~\cite{CeAl80, PeWa:92}, PBE~\cite{PeBu96},
the hybrid functionals HSE06~\cite{HSE:06}, B3LYP~\cite{Becke:93, LeYa88, B3LYP}, PBE0~\cite{PBE0:96}, and in the Hartree-Fock approximation.
For the adsorption and STM simulations, we used a Si$(111)$-$(6 \times 6)$ slab at the PBE lattice constant of Si ($\sim 5.47~\AA$)
with two passivating H layers, $20~\AA$~vacuum region,
a $2\times2\times1$ $\vec{k}$-point grid including the $\Gamma$~point and the PBE functional for exchange and correlation.
Van-der-Waals interactions were treated within the Grimme D2~\cite{Grimme:06, Barone:09} and D3~\cite{Grimme-D3:10} approaches
and compared to the Tkatchenko-Scheffler (TS) method~\cite{Tkatchenko:09},
in which the free-atom $C_6$ coefficients, polarizabilities, and vdW radii are rescaled by a Hirshfeld partitioning of the electron density.
We also explored the more recent many-body dispersion (MBD) technique~\cite{Tkatchenko:12},
in which the atomic response functions are represented by quantum harmonic oscillators
and the screened long-range many-body vdW energy
is computed from the adiabatic connection fluctuation dissipation theorem within the dipole approximation.
Constant-current STM images at a bias voltage~$V$ were simulated subsequently in the spirit of Tersoff and Hamann~\cite{tersoff:83}
as $\varrho_0$-isosurfaces of the integrated local density of states (ILDOS)~\cite{Geisler:12, Suzuki:13, Geisler:13, Geisler-TMs:15, GeislerPentcheva-LNOLAO-Resonances:18}:
$\varrho(\vec{r}) = \int_{0}^{eV} \vert \text{d} \varepsilon \vert \, \sum_{n \vec{k}} \vert \psi_{n \vec{k}} (\vec{r}) \vert^2 \ \delta(\varepsilon - \varepsilon_{n \vec{k}} + E_{\text{F}})$,
$\left\lbrace z(x,y) = z \, : \, \varrho(x,y,z) = \varrho_0 \right\rbrace$; we used $\varrho_0 = 10^{-7}/(\text{a.u.})^3$.
SP~STM images correspond to differences $z_{\uparrow} - z_{\downarrow}$ of two STM images derived from the individual spin channels.

% (and MnPc) 
The DFT treatment of FePc is nontrivial due to numerical instabilities.
Calculations employing the PBE exchange-correlation functional often lead to a symmetry-broken $D_{2h}$ ground state,
whereas hybrid functionals such as B3LYP preserve the $D_{4h}$ symmetry of the molecule~\cite{MaromKronik-FePcMnPc:09}.
% Moreover, the actual ground state of FePc, for instance, is under debate~\cite{Gruyters:12}.
We used the B3LYP electronic structure as reference,
which we found to be qualitatively reproduced by PBE$+U$~\cite{QE-LDA-U:05} with a small $U_\text{Fe}=3$~eV,
as is has also been utilized in the literature~\cite{Mugarza:12, Herper-FePc-Co001:14}.
In order to obtain unbiased binding energies, we did not apply a Hubbard~$U$ in those cases,
assuring carefully that the symmetry is not broken.

\begin{figure*}[t]
	\centering
	\includegraphics{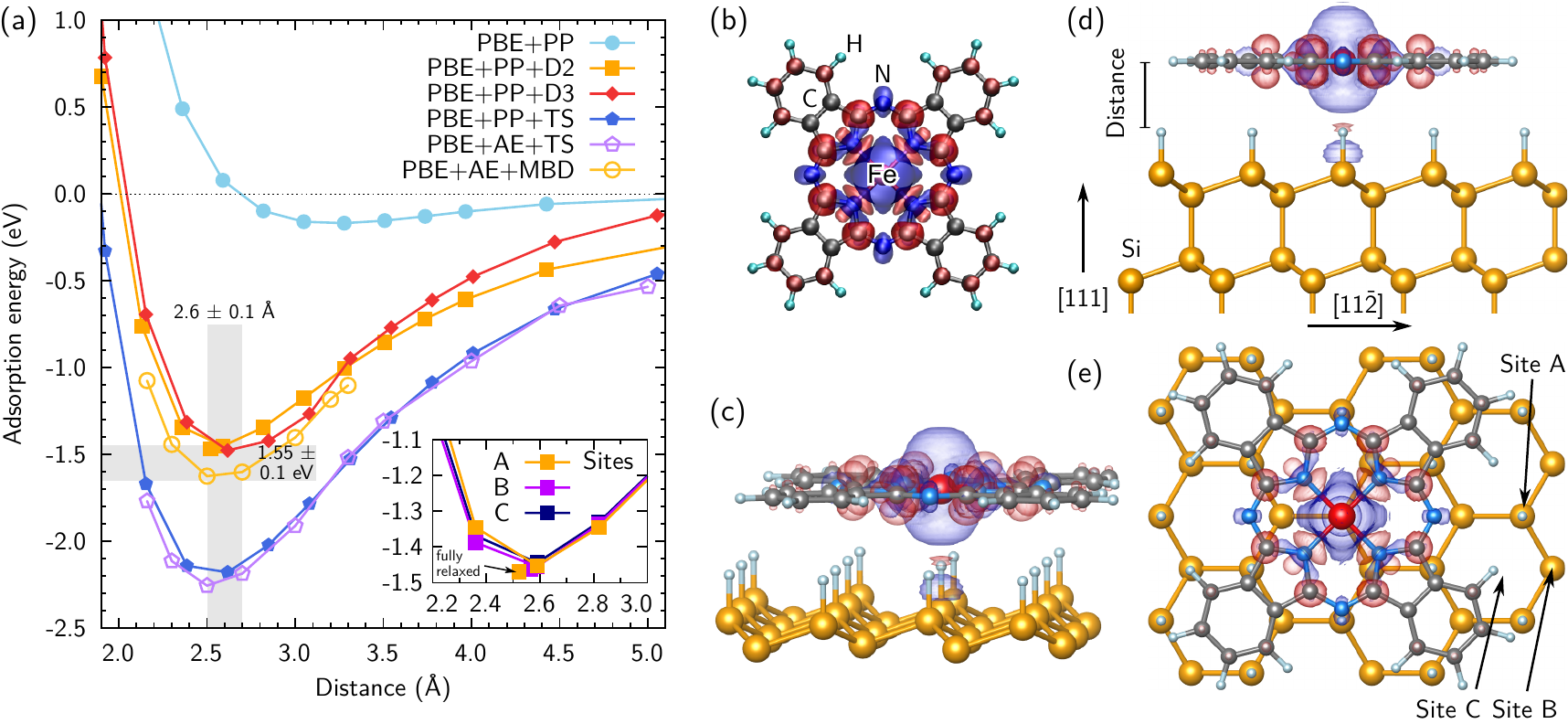}
	\caption{Adsorption of a single FePc molecule on the H/Si$(111)$ surface. (a)~Adsorption energies $E_{\text{Ads}}$ as functions of the molecule-surface distance, comparing different vdW methods. The inset compares three distinct adsorption sites. The curves have been obtained from rigid atomic structures, whereas the unconnected data points correspond to fully optimized atomic positions (see text). (b)~Optimized gas-phase FePc geometry and corresponding spin density. (c)~Three-dimensional perspective view, (d)~side view, and (e)~top view of an FePc molecule adsorbed at a H/Si$(111)$~A site as obtained within the PBE$+$PP$+$D2 framework ($U_\text{Fe}=3$~eV). The spin densities exhibit a polarized H-Si bond below the Fe ion (blue/red corresponding to positive/negative).}
	\label{fig:FePc-on-HSi}
\end{figure*}

\section{\boldmath Adsorption of F\lowercase{e}P\lowercase{c} on H/S\lowercase{i}$(111)$}

Figure~\ref{fig:FePc-on-HSi}(a) shows the adsorption energies
of a single FePc molecule on the H/Si$(111)$ surface as functions of the molecule-surface distance
calculated from DFT total energies:
\begin{equation*}
E_{\text{Ads}} = E_{\text{Molecule on surface}}^{} - E_{\text{Isolated molecule}}^{} - E_{\text{Ideal surface}}^{}
\text{.}
\end{equation*}
The adsorption energy curves exhibit a typical potential shape, diverging for small distances
and approaching zero (corresponding to the total energy sum of separated systems) from below for large distances.
Three distinct adsorption sites are compared:
centering Fe above the first Si layer positions, i.e., above one of the H atoms (A),
above the second/third Si layer positions (B),
and above the fourth Si layer positions (C),
which correspond to the hollow centers of the topmost bilayer rings [Fig.~\ref{fig:FePc-on-HSi}(e)].
The curves have been obtained from rigid atomic structures, i.e., no relaxation beyond the individual equilibrium geometries
of molecule and surface
has been performed.
Subsequent calculations optimizing all atomic positions at the adsorption distance have shown that adsorption at site~A is most stable [Fig.~\ref{fig:FePc-on-HSi}(a)].
Energy differences with respect to sites~B and~C amount to only a few meV;
the same applies to different rotation angles of the molecule.
Thus, we conclude that FePc is highly mobile on the H/Si$(111)$ surface.

The resulting adsorption energies and distances are summarized in Table~\ref{tab:FePc-Distance}.
Interestingly, already PBE without further vdW corrections provides a shallow adsorption minimum of $\sim 0.18$~eV near $3.2~\AA$
despite the chemical inertness~\cite{Neuwald-HSi111-Inertness:92} of the H/Si$(111)$ surface.
However, this adsorption energy is certainly far too low, and the equilibrium molecule-surface distance will be severely overestimated.
Indeed, we find that FePc attaches more closely to the substrate if vdW interactions are taken into account;
all considered vdW methods provide adsorption distances within an interval of $2.6 \pm 0.1~\AA$.
The adsorption energies are more strongly spread:
Predictions based on the D2, D3, and MBD approaches range from $1.45$ to $1.65$~eV,
whereas the TS method results in a much stronger binding of about $2.2$ to $2.27$~eV.
Thus, a many-body treatment of the vdW interactions~\cite{Tkatchenko:12} corroborates the results of D2 (and D3).
Interestingly, we find very good agreement for two distinct implementations (PP vs.\ AE)
of the TS method in two individual codes [Fig.~\ref{fig:FePc-on-HSi}(a)].

The results shown in Fig.~\ref{fig:FePc-on-HSi}(a) and Table~\ref{tab:FePc-Distance}
can be compared
to other hybrid inorganic-organic systems.
For 3,4,9,10-perylene tetracarboxylic dianhydride (PTCDA) on Ag$(111)$~\cite{Ruiz:12},
PBE provides a shallow adsorption minimum of about $0.3$~eV
at a distance $\sim 1~\AA$ larger than the experimental value.
Inclusion of D2 vdW corrections severely overestimates the adsorption energy by $\approx 1$~eV ($40\,\%$),
simultaneously providing an almost correct molecule-surface distance.
In that case, screening effects of the metal substrate play an important role,
and the performance of D2 will be significantly better for the present system.
For anthracene and pentacene on Ag$(111)$~\cite{Morbec-AnthraPentaAg111:17},
PBE predicts a shallow adsorption minimum of $\sim 0.1$~eV,
and the surface-molecule distance is lowered by about $1~\AA$ after inclusion of vdW interactions.
The adsorption energy obtained with the MBD approach is in almost perfect agreement with the experimental value.

\begin{table}[b]
	\centering
%	\vspace{-1.5ex}
	\caption{\label{tab:FePc-Distance}Optimized distances and adsorption energies as obtained by different vdW approaches for a single FePc molecule adsorbed on the H/Si$(111)$ surface at site~A (cf.~Fig.~\ref{fig:FePc-on-HSi}).}
	\begin{ruledtabular}
	\begin{tabular}{lcccccc}
 & \multicolumn{4}{c}{PP} & \multicolumn{2}{c}{AE}	\\
\cline{2-5}
\cline{6-7}
 & PBE & $+$D2 & $+$D3 & $+$TS & $+$TS & $+$MBD	\\
\hline
Distance (\AA)						& $3.2$  & $2.52$ & $2.7$ & $2.55$ & $2.55$ & $2.58$	\\
$\vert E_{\text{Ads}} \vert$ (eV)		& $0.18$ & $1.45$ & $1.5$ & $2.2$  & $2.27$ & $1.65$	\\
	\end{tabular}
	\end{ruledtabular}
\end{table}

For the related case of vdW-bound 2D materials,
recent comparative work revealed that the TS method predicts the adsorption energy with least precision,
whereas D2, D3, and in particular a many-body treatment of the vdW interactions considerably improve the accuracy~\cite{Tawfik:18}.
In conjunction with the analysis above, we therefore conclude a tentative adsorption energy of $1.55 \pm 0.1$~eV for FePc on H/Si$(111)$.

The $D_{4h}$ environment of the Fe$^{2+}$ ion results in a FePc magnetic moment of $2~\mu_{\text{B}}$ as expected from ligand field theory.
Its spin density is shown in Fig.~\ref{fig:FePc-on-HSi}(b).
The adsorbed molecule retains its gas-phase magnetic moment for all methodologies considered here,
and it becomes obvious from Figs.~\ref{fig:FePc-on-HSi}(c)-(e) that also the shape of the spin density is largely preserved,
apart from a slight compression of the lobe directed towards the substrate.
Interestingly, we observe a sizable spin polarization of the H-Si bond in the substrate below the Fe ion.

\begin{figure}[t]
	\centering
	\includegraphics{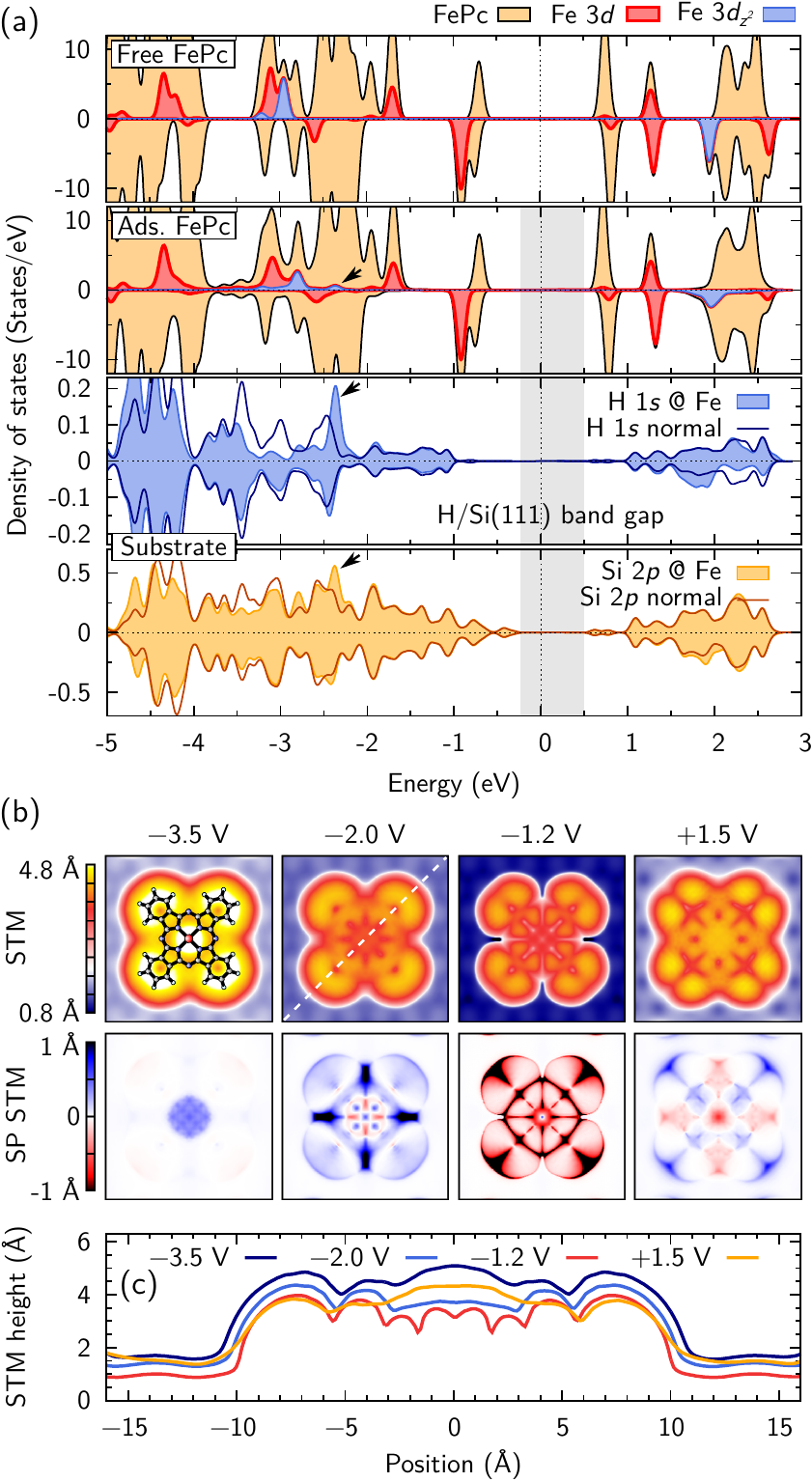}
	\caption{Electronic structure of a single FePc molecule adsorbed at a H/Si$(111)$ A~site as obtained within the PBE$+$PP$+$D2 framework ($U_\text{Fe}=3$~eV). (a)~Density of states of gas-phase FePc, adsorbed FePc, and of the H and Si atoms in the (sub-)surface layer at the adsorption site. The black arrows mark an Fe-$3d_{z^2}$-H-$1s$-Si-$2p$ resonance, and the grey area depicts the band gap of the substrate. (b)~Simulated STM and SP~STM images together with a structural model of FePc. The bias voltages correspond to the panels in~(a). (c)~Line scans along the diagonal of the STM panels in~(b) for different bias voltages. The height is measured with respect to the FePc position above the substrate, and the molecule has an actual diameter of about $15~\AA$.}
	\label{fig:STM}
\end{figure}

At first sight,
it appears that the proximity of the surface has only modest impact on the electronic structure of the molecule [Fig.~\ref{fig:STM}(a)],
which is different e.g.\ for anthracene and pentacene on Ag$(111)$~\cite{Morbec-AnthraPentaAg111:17}.
Relative to gas-phase FePc,
the band gap between the highest occupied molecular orbital (HOMO) and the lowest unoccupied molecular orbital (LUMO) is almost unaffected.
Both reside within the valence band or the conduction band of the substrate, respectively.
A finite density of states emerges at $-3.5$~eV.
Specifically, the Fe~$3d_{z^2}$ states are strongly broadened;
while the occupied majority-spin state at $-3$~eV shifts to $-2.8$~eV, its unoccupied equivalent in the minority-spin channel retains its position.
The H~$1s$ and Si~$2p$ states below the Fe ion clearly respond to the presence of the molecule as well,
reflecting the induced spin polarization
and exhibiting an Fe-$3d_{z^2}$-H-$1s$-Si-$2p$ resonance [Fig.~\ref{fig:STM}(a)].
Close inspection of the atomic positions reveals that Fe and H leave their respective planes and reduce their distance by $0.05~\AA$.
This attractive interaction shows that the system enters a mixed covalent-vdW-bonding state,
which exemplifies that the inclusion of vdW interactions does not only \textit{quantitatively} improve the results,
but is of \textit{qualitative} importance for a proper modeling of FePc adsorption on H/Si$(111)$.
Hence, this system emerges as interesting test case
for different vdW corrections in the field of electronic structure theory~\cite{Grimme:06, Grimme-D3:10, Tkatchenko:09, Tkatchenko:12}.
Specifically, exploring the performance of 'true' vdW density functionals is of importance.~\cite{LangrethLundqvist:04, Perez-Soler-vdw:09, LeeLundqvistLangreth-vdW-DF2:10}.

Figure~\ref{fig:STM}(b) shows simulated STM and SP~STM images for an adsorbed single FePc molecule on the H/Si$(111)$ surface at site~A,
together with a line scan for different bias voltages [Fig.~\ref{fig:STM}(c)].
For the filled-state images ($V<0$), the H sites of the substrate can be identified by white spots~\cite{Geisler-TMs:15}.
The $D_{4h}$ symmetry of the molecule is clearly reflected in all panels.
The distinct lobes that can be seen for $-1.2$~V (cross-like shape) merge for voltages of higher magnitude (flower-like shape),
consistent with experiment~\cite{Gruyters:12}.
Moreover, while for $-1.2$~V the outer parts of the lobes are dominant,
a feature appears in the center of the molecule with increasing bias voltage
that can be clearly seen from the line scans
and is also observed experimentally~\cite{Gruyters:12}.
In our simulations, the molecule is detected $\sim 4$-$5~\AA$ above its physical position
and $\sim 3$-$4~\AA$ above the substrate-related background,
with an apparent diameter of $21 \pm 1~\AA$ (structurally $15~\AA$).
Experimentally, its signature appears a little larger,
exhibiting a height of $\sim 5~\AA$ above the substrate-related background
and a diameter of $\sim 25~\AA$
at $-2.25$ to $-2.65$~V and $10$-$20$~pA~\cite{Gruyters:12}.
Overall, the agreement between simulations and experiment is very good.

We find that the SP~STM magnetic contrast is significant, exceeding $\pm 1~\AA$ for $-2$ and $+1.5$~V.
Since SP~STM does not correspond to a mapping of the spin density (Fig.~\ref{fig:FePc-on-HSi}),
but of the states near the Fermi energy,
we see a negative response for $-1.2$~V (red, largest at the outer lobe parts) that switches to a positive signal for $-2$~V (blue, largest between the lobes).
Finally, inclusion of the majority-spin Fe~$3d_{z^2}$ state for a bias voltage of $-3.5$~V leads to a positive feature in the center.
In turn, this feature appears negative and even more pronounced for a mapping of the unoccupied states beyond $+2$~V (not shown).

\section{Binding energies of transition metal ions in P\lowercase{c} molecules}

Experiments have suggested that FePc molecules can release their Fe center upon room-temperature adsorption on H/Si$(111)$,
thereby leading to a subsurface doping of H/Si$(111)$ by Fe~\cite{Gruyters:13}.
In order to shed light on the underlying mechanism, it is important to understand the energetics of TM bonding in the Pc ring.
To give additional credibility to our results, we carried out calculations for various gas-phase Pc molecules
not only with Fe, but also with other $3d$ TM atoms and with Si as center,
and compare the values obtained with a variety of exchange-correlation functionals.
This information will further be used in the next section to extract DFT total energy differences of distinct reactions at the H/Si$(111)$ surface.

% In order to shed light on the mechanism behind the suggested subsurface doping based on FePc deposition~\cite{Gruyters:13},
% we start by exploring energetic aspects of different gas-phase Pc molecules, particularly FePc.
% This information will be used in the next section to extract DFT total energy differences of distinct reactions at the H/Si$(111)$ surface.
% accounting also for a possible influence of the methodology (PP vs.\ AE) and of the exchange-correlation functional.

\begin{figure}[t]
	\centering
	\includegraphics{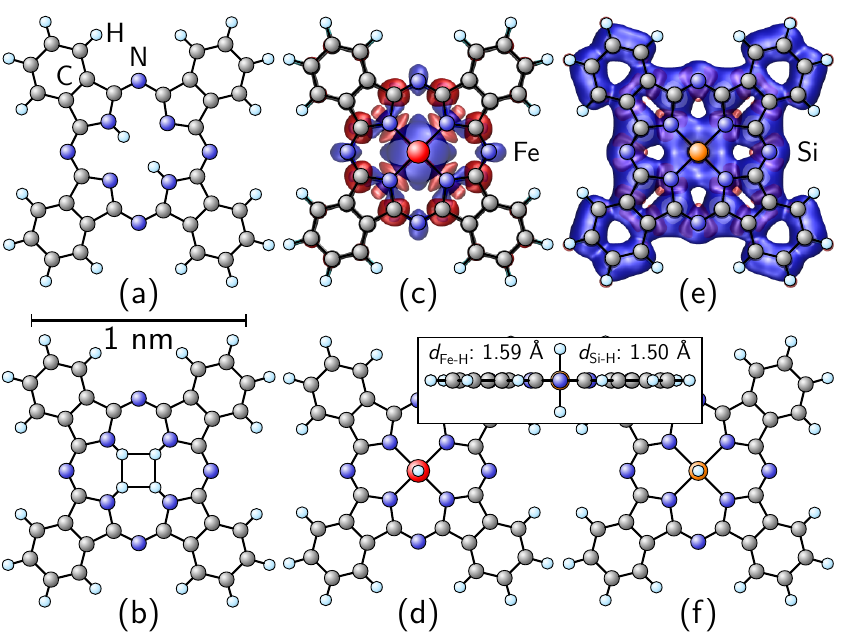}
	\caption{Optimized structures of Pc rings with different centers. The empty Pc ring can bind two H atoms in different configurations, the realization shown in~(a) being the most stable; four H atoms~(b); a single Fe atom~(c); or a single Si atom~(e). Moreover, the latter two can bind two additional H atoms above and below the molecular plane (d), (f). The corresponding binding energies are given in Tables~\ref{tab:Pc-BindingEnergies} and~\ref{tab:FePc-BindingEnergies}. For the cases of of FePc and SiPc, spin densities are shown (blue/red corresponding to positive/negative)}
	\label{fig:MolecularStructures}
\end{figure}

Figure~\ref{fig:MolecularStructures} shows optimized atomic structures of selected Pc molecules.
The corresponding binding energies of different central atoms in an empty Pc ring are given in Table~\ref{tab:Pc-BindingEnergies}.
They are defined as
\begin{equation*}
\label{eq:BindingEnergy}
E_{\text{Bind}} = E_{\text{Molecule}}^{} - E_{\text{Empty Pc ring}}^{} - \sum_i \Delta N_{i}^{} \, E_{i}^{\text{Atom}}
\text{,}
\end{equation*}
where the first two terms are the total energies of the molecule and the empty Pc ring as reference,
$\Delta N_{i}^{}$ is the difference of atoms of species~$i$ between molecule and reference,
and $E_{i}^{\text{Atom}}$ denotes the total energy of an isolated atom of species~$i$.
Two H atoms are bound twice as strong in the Pc ring ($9.22$~eV) as in a H$_2$ molecule (PBE: $4.45$~eV), % for two H atoms, i.e., H2 has not been not divided by two!
while four bound H atoms are less favorable than the former case plus an isolated H$_2$ molecule.
Most notably, we find that Fe is even more strongly bound ($9.59$~eV).
In order to see if this surprisingly high Fe binding energy is an exception,
we also studied other $3d$ TM centers in the Pc ring
and found similar results, Cu$/$Ti exhibiting the lowest$/$highest binding energy ($7.69/10.87$~eV).
To account for possible Si exchange processes between substrate and molecule,
we also simulated a Si atom in a Pc ring, and found it to be bound with $8.51$~eV, a value almost twice as high as in Si bulk (PBE: $4.6$~eV).
Both FePc and SiPc can bind two additional H atoms above and below the molecular plane.
While in the case of FePc the formation of an isolated H$_2$ molecule is energetically favored,
SiPc-2H is stable against decomposition with an energy gain of $1.61$~eV.

\begin{table}[b]
	\centering
	\vspace{-1.5ex}
	\caption{\label{tab:Pc-BindingEnergies}Binding energies and magnetic moments of different central atoms in an empty Pc ring, calculated within PBE$+$PP. 2H, 4H, Fe $+$ 2H, and Si $+$ 2H correspond to (a), (b), (d), and (f) in Fig.~\ref{fig:MolecularStructures}, whereas Fe and Si correspond to (c) and (e), respectively.}
	\begin{ruledtabular}
	\begin{tabular}{lcccccc}
 & Fe & Fe $+$ 2H & Si & Si $+$ 2H & 2H & 4H	\\
\hline
$\vert E_{\text{Bind}} \vert$~(eV)	& $9.59$ & $12.27$ & $8.51$ & $14.57$ & $9.22$ & $12.58$	\\
$m$~($\mu_{\text{B}}$)				& $2$ & $0$ & $2$ & $0$ & $0$ & $2$	\\
	\end{tabular}
	
	\begin{tabular}{lcccccc}
 & Ti & Cr & Mn & Fe & Co & Cu	\\
\hline
$\vert E_{\text{Bind}} \vert$~(eV)	& $10.87$ & $9.52$ & $9.11$ & $9.59$ & $9.96$ & $7.69$	\\
$m$~($\mu_{\text{B}}$)				& $2$ & $4$ & $3$ & $2$ & $1$ & $1$	\\
	\end{tabular}
	\end{ruledtabular}
\end{table}

\begin{table}[b]
	\centering
%	\vspace{-1.5ex}
	\caption{\label{tab:FePc-BindingEnergies}Binding energies $\vert E_{\text{Bind}} \vert$~(eV) of a central Fe atom or two H atoms in FePc and Pc, respectively, calculated with different exchange-correlation functionals and Hartree-Fock (HF). For PBE, benchmarking with AE results underlines the reliability of the PP method.}
	\begin{ruledtabular}
	\begin{tabular}{lccccccc}
 & PP & \multicolumn{6}{c}{AE}	\\
\cline{2-2}
\cline{3-8}
 & PBE & PBE & LDA & B3LYP & PBE0 & HSE06 & HF	\\
\hline
Fe in FePc	& $9.59$ & $9.57$ & $11.74$ & $8.23$ & $8.69$ & $8.72$ & $5.81$	\\
2H in Pc	& $9.22$ & $9.24$ & $10.59$ & $9.15$ & $9.24$ & $9.27$ & $6.27$	\\
	\end{tabular}
	\end{ruledtabular}
\end{table}

In addition to the binding energies, Table~\ref{tab:Pc-BindingEnergies} also lists the spin magnetic moments of different molecules.
The values for the more common types such as MnPc, FePc, CoPc, or CuPc are well known~\cite{KondoMnPc:07, Mugarza:12}.
We also find (high) magnetic moments for TiPc and, particularly, CrPc.
Most peculiarly, we observe a very stable magnetic moment of $2~\mu_{\text{B}}$ for SiPc,
the spin density extending over the entire molecule [Fig.~\ref{fig:MolecularStructures}(e)],
while that of FePc is confined to the center.

In order to further validate the very high FePc binding energy we reported above,
Table~\ref{tab:FePc-BindingEnergies} compares results calculated with different methodologies (PP vs.\ AE) and exchange-correlation functionals
for FePc and a Pc ring with two H atoms [Fig.~\ref{fig:MolecularStructures}(a)].
In particular, hybrid functionals like B3LYP, PBE0, or HSE06 are commonly used for organic molecules~\cite{FePc-Schweden:06, HSE:03, HSE:06b}.
% The results of Hartree-Fock calculations are provided for completeness.
%
The PBE results obtained within the PP and AE frameworks are in excellent agreement.
LDA shows the well-known overbinding.
For FePc, the binding energies provided by the hybrid functionals are roughly $1$~eV smaller than the PBE results, but still very high.
For the Pc ring with two H atoms, hybrid functional results are similar to PBE.
Hartree-Fock calculations lead to significantly smaller binding energies for both systems,
but since they ignore all correlation effects, their appropriateness for larger organic molecules can be questioned.
The magnetic moment of FePc is $2~\mu_{\text{B}}$ in all cases.
While Fe is more strongly bound than two H atoms in LDA and PBE, the reversed prediction is made by hybrid functionals and Hartree-Fock.
HSE06 and PBE0 provide very similar results, which shows that the screening of the Coulomb interaction in HSE06 is irrelevant on the present length scale.

On the basis of these first-principles findings we conclude that the very high binding energy we obtained for a Fe center in a Pc ring is indeed reasonable.

\section{Subsurface doping}

After this excursion, we now return to FePc and possible mechanisms to achieve subsurface Fe doping
by room-temperature deposition of FePc on H/Si$(111)$,
as it has been proposed recently on the basis of STM imaging~\cite{Gruyters:13}.
The comprehensive DFT studies presented in the previous section, which are summarized in Tables~\ref{tab:Pc-BindingEnergies} and~\ref{tab:FePc-BindingEnergies},
show that Fe is very strongly bound to the organic part of the molecule and thus quite difficult to extract.
This raises the question whether the suggested doping strategy is realistic,
particularly since a combined experimental-theoretical analysis of the STM observations remained inconclusive~\cite{Gruyters:13, Geisler-TMs:15}.

\begin{table}[b]
	\centering
%	\vspace{-1.5ex}
	\caption{\label{tab:TM-in-Si-FePc-Doping}Different reactions for the injection of isolated interstitial or substitutional Fe impurities into the H/Si$(111)$ subsurface layers [cf.~Fig.~\ref{fig:DopingScenarios}] from FePc and FePc-2H. $\Delta E$ is the DFT total energy difference between the final and the initial state; a negative value indicates that the final configuration is preferred.}
	\begin{ruledtabular}
	\begin{tabular}{lc}
Reaction & $\Delta E$ (eV)	\\
\hline
FePc $+$ pure H/Si$(111)$ $\to$ &	\\
\hline
Fe subs.\ $+$ SiPc											&	$+2.79$		\\
Fe subs.\ $+$ SiPc-2H $+$ $2\times$ H-Si bond broken			&	$+3.51$		\\
Fe subs.\ complex with Si self-inter.\ $+$ empty Pc ring	&	$+8.60$		\\
Fe subs.\ $+$ Si self-inter.\ $+$ empty Pc ring				&	$+10.73$		\\
Fe inter.\ $+$ empty Pc ring									&	$+6.15$		\\
Fe inter.\ $+$ Pc $+$ $2\times$ H-Si bond broken				&	$+3.70$		\\
\hline
FePc-2H $+$ pure H/Si$(111)$ $\to$ &	\\
\hline
Fe subs.\ $+$ SiPc-2H										&	$-0.60$		\\
Fe subs.\ complex with Si self-inter.\ $+$ Pc				&	$+2.05$		\\
Fe inter.\ $+$ Pc											&	$-0.40$		\\
	\end{tabular}
	\end{ruledtabular}
\end{table}

In a first step towards addressing this complex question,
DFT total energy differences~$\Delta E = E_{\text{Final}} - E_{\text{Initial}}$ of several distinct reactions (Table~\ref{tab:TM-in-Si-FePc-Doping})
can be considered under the constraint that the number of atoms of each species has to be conserved.
Positive~$\Delta E$ indicates an endothermic reaction.
For instance, the Fe atom could leave the FePc molecule and replace a Si atom in the host (substitutional Fe);
subsequently, the Si atom is incorporated in the rest of the molecule, forming SiPc
in which the Si atom is more strongly bound than in bulk Si, as discussed above, leading to $\Delta E = +2.79$~eV for this scenario.
Additionally, two H atoms could leave the passivating H layer and adsorb on the SiPc molecule [Fig.~\ref{fig:MolecularStructures}(f)].
The defect energy associated with a missing H atom can be approximated by the H-Si binding energy (PBE: $3.39$~eV, $\Delta E = +3.51$~eV).
A third possibility is the formation of a Si self-interstitial impurity,
which can either form a complex with the substitutional Fe impurity or be isolated from it.
One can see from Table~\ref{tab:TM-in-Si-FePc-Doping} that all scenarios based on a removal of Fe from FePc
with subsequent implantation into the H/Si$(111)$ matrix are energetically highly unfavorable,
even when compared to typical room-temperature energies ($25$~meV)
and irrespective of the final Fe site, i.e.,  interstitial or substitutional.
The situation remains unchanged if one tentatively lowers the PBE binding energy of FePc to a hybrid functional result (Table~\ref{tab:FePc-BindingEnergies}).
In addition to the total energies of different initial and final states,
kinetic barriers caused by transition states might inhibit the Fe exchange process.

\begin{figure}[t]
	\centering
	\includegraphics{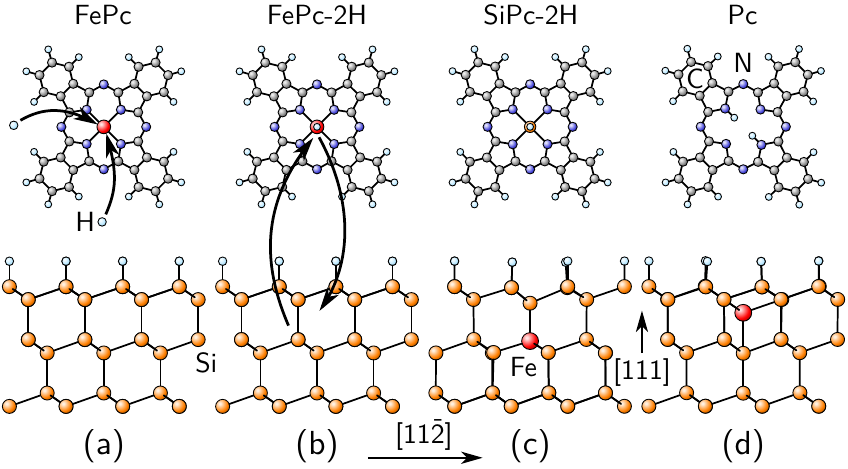}
	\caption{Illustration (not to scale) of different reactions with negative $\Delta E$ for the injection of interstitial or substitutional Fe impurities into the H/Si$(111)$ subsurface layers from FePc. Adsorption of two excess H atoms on FePc~(a) leads to FePc-2H~(b). Subsequently, the Fe atom can interchange with a Si atom~(c) or move to an interstitial site~(d), while the two H atoms stabilize the molecule.}
	\label{fig:DopingScenarios}
\end{figure}

The origin of the highly positive values of $\Delta E$ lies in the stability of the initial state, but also in the unfavorable final states
considered so far.
Let us now assume that two H atoms adsorb on a gas-phase FePc molecule [Figs.~\ref{fig:MolecularStructures}(d) and~\ref{fig:DopingScenarios}(a)].
If the Fe atom occupies a substitutional site after the reaction [Fig.~\ref{fig:DopingScenarios}(c)],
the removed Si atom can be incorporated in the Pc ring, which results in SiPc with two additional H atoms above and below the molecular plane (SiPc-2H),
which is very stable (Table~\ref{tab:Pc-BindingEnergies}).
This scenario optimizes both $E_{\text{Initial}}$ and $E_{\text{Final}}$
and leads to $\Delta E < 0$.
If the Fe atom is integrated on an interstitial site [Fig.~\ref{fig:DopingScenarios}(d)],
not SiPc-2H, but simply Pc remains after the reaction,
resulting in $\Delta E < 0$ as well.

Hence, the latter two mechanisms are statistically relevant (neglecting kinetic barriers).
The crucial point is the adsorption of H atoms on the FePc molecule prior to the reaction at the H/Si$(111)$ surface.
This can occur if, in addition to the FePc molecules, H atoms are evaporated from the crucible in the experiment,
or if excess H atoms exist in the vicinity of the passivated surface.
Moreover, a destabilization of FePc due to interactions with the H atoms in the passivating layer, as discussed above, facilitates the process.

We conclude from this analysis that subsurface doping of H/Si$(111)$ by room-temperature deposition of FePc molecules is possible,
but the probability of the necessary steps in the reaction is low due to strongly bound Fe ion in FePc.
More detailed investigations should be done in this field.

\section{Summary}

We studied the adsorption of iron phthalocyanine (FePc) on the passivated H/Si$(111)$ surface from first principles.
According to our findings, the organic molecule is predominantly physisorbed
with a distance to the surface of $2.6 \pm 0.1~\AA$ and an adsorption energy of $1.55 \pm 0.1$~eV,
but also shows significant resonance with the underlying substrate.
We used this paradigmatic case of a system in a mixed covalent-van-der-Waals-bonding state to compare the impact of different approaches to van-der-Waals interactions.
(Spin-polarized) scanning tunneling microscopy (SP~STM) images were simulated,
selectively accessing different molecular orbitals via the applied bias voltage, similar to scanning tunneling spectroscopy.
Comparison with experimental STM images revealed very good agreement.
We found a considerable magnetic contrast exceeding $\pm 1~\AA$ in the SP~STM images for $-2$ and $+1.5$~V.
% Binding energies of different (transition metal) atoms in the center of the Pc ring were presented,
% showing in particular that Fe is strongly bound in the molecule (about $9.6$~eV).
% Finally, we discussed different mechanisms for subsurface doping by room-temperature FePc deposition and pointed out two feasible reactions.

In order to provide first insight into the mechanism behind a suggested subsurface doping strategy that involves room-temperature deposition of FePc on H/Si$(111)$,
we presented binding energies of different (transition metal) atoms in the center of the Pc ring,
which particularly revealed that Fe is strongly bound (about $9.6$~eV).
This surprising result was confirmed by different methodologies, including hybrid functionals for exchange and correlation, and holds for other $3d$ transition metal centers as well.
Despite this finding, we identified two feasible (exothermic) reactions by analyzing several distinct mechanisms to achieve subsurface Fe doping.
Concomitantly, we highlighted the crucial role of a preceding destabilization of FePc, for instance, by pre-adsorbed H atoms,
which subsequently strongly stabilize the final state of the reaction.

\section{Acknowledgments}

%\begin{acknowledgments}

Computing time was granted by the Center for Computational Sciences and Simulation of the University of Duisburg-Essen.

%\end{acknowledgments}

% \bibliography{BibTeX/Dissertation,BibTeX/dft,BibTeX/MnSi,BibTeX/FeSi,BibTeX/GaAs,BibTeX/TM-in-Si,BibTeX/Heusler,BibTeX/ZnO,BibTeX/Oxides}


\end{document}